

\input phyzzx.tex
\overfullrule0pt

\def\ie{{\it i.e.}}

\def\bold#1{\setbox0=\hbox{$#1$}%
     \kern-.025em\copy0\kern-\wd0
     \kern.05em\copy0\kern-\wd0
     \kern-.025em\raise.0433em\box0 }
\Pubnum={VAND-TH-94-1}
\date={January 1994}
\pubtype{}
\titlepage

\vskip1cm
\title{ \bf Decays of a fermiophobic Higgs}
\author{Marco A. D\'\i az and Thomas J. Weiler}
\vskip .1in
\centerline{Department of Physics and Astronomy}
\centerline{Vanderbilt University, Nashville, TN 37235}
\vskip .2in

\centerline{\bf Abstract}
\vskip .1in

We explore the phenomenology of a fermiophobic Higgs: a Higgs
whose couplings to fermions are suppressed. We calculate the branching
ratios of a Higgs decaying to $\gamma\gamma$, $W^*W^*$, $Z^*Z^*$,
$Zb\bar b$, $Z\gamma$, $\gamma b\bar b$, and final states involving
vector mesons like $\Upsilon$, $J/\Psi$ and $\rho$. In order to
calculate these branching ratios we perform a complete one-loop
renormalization of the vertices $HZ\gamma$ and $H\gamma\gamma$.
The decay mode $H\rightarrow \gamma\gamma$ is near unity for
a Higgs below the $W$ mass, which provides a clean way of discovering
a light fermiophobic Higgs. Interesting modes involving the vector mesons
$Z$, $\gamma$, $\rho$, $J/\Psi$, and $\Upsilon$ are carefully analyzed.

\vskip 2.cm
* Presented by Marco A. D\'\i az at the Third Annual SAHEP Gulf Shores
Meeting, Gulf Shores, Alabama, January 3-5, 1994.
\vfill

\endpage

\voffset=-0.2cm

\REF\fermphob{T.J. Weiler, VIII Vanderbilt International Conference
on High Energy Physics, ed. J. Brau and R.S. Panvini, 1987, World
Scientific Publications; H. Pois, T.J. Weiler, and T.-C. Yuan,
{\it Phys. Rev. D} {\bf 47}, 3886 (1993).}

\REF\SMarW{A. Stange, W. Marciano, and S. Willenbrock, report
number FERMILAB-PUB-93/142-T.}

It is perfectly possible that the Higgs mechanism responsible for
generating the masses of the gauge bosons is independent from the
mechanism that generates the fermion masses. In this way, the existence
of a ``fermiophobic'' Higgs is possible\refmark\fermphob, whose tree
level couplings to fermions are zero\foot{This kind of Higgs
has been also analyzed in ref.~[\SMarW].}.

\REF\BDHR{V. Barger, N.G. Deshpande, J.L. Hewett, and T.G. Rizzo,
report number ANL-HEP-PR-92-102.}

Higgs fields in representations other than doublets
contribute to gauge boson
masses but do not couple directly to fermions. Nevertheless, the $\rho$
parameter restricts the vacuum expectation values of those
Higgs fields. But even in models with only Higgs doublets we can find
fermiophobia\refmark\BDHR.
For example in two Higgs doublets model of type I, in
which only one Higgs doublet (say, $\Phi_1$) couples to fermions, the
second Higgs field $\Phi_2$ will have zero couplings to fermions. One
way to implement this fermiophobia is imposing the discrete symmetry
$\Phi_2\rightarrow-\Phi_2$. But this fermiophobia will be partial,
since the mass eigenstates ($H$ and $h$) are a mixture of $\Phi_1$
and $\Phi_2$. If we fine-tune the mixing angle $\alpha$ to $\sin\alpha=0$
then one of the Higgs mass states (say $H$) will be fermiophobic at
tree level.

\FIG\Hfbarf{Fynman diagrams contributing to the one-loop renormalization
of the vertex $Hf\bar f$ in the case of a fermiophobic Higgs $H_F$. The
sum of the diagrams is infinite, and a counterterm is necessary.}

The fine-tuning $\sin\alpha=0$ at tree level
does not guarantee perfect fermiophobia.
There is a one-loop induced vertex $Hf\bar f$.
In Fig.~\Hfbarf\ we show the one-loop
contributions to the vertex $Hf\bar f$ in the fermiophobic case. The
sum of these graphs is infinite; therefore the vertex $Hf\bar f$ is
infinitely renormalizable, and we need an experimental measurement in
the Higgs sector to fix the counterterm. This arbitrariness leads us to
defene an extreme fermiophobic Higgs $H_F$, as that obtained when we
set the renormalized vertex $Hf\bar f$ to zero. We stress that
a more realistic Higgs would be partially fermiophobic, \ie\ with
supressed but non-zero couplings to fermions.

\REF\diazweiler{Marco A. D\'\i az and Thomas J. Weiler, in preparation.}
\REF\BarrPR{These vertices are renormalized when one photon is
on-shell in:
A. Barroso, J. Pulido, and J.C. Rom\~ ao, {\it Nucl.
Phys.} {\bf B267}, 509 (1986), and J.C. Rom\~ao and A. Barroso, {\it Nucl.
Phys.} {\bf B272}, 693 (1986).}

In order to calculate the different branching ratios of interest,
we perform a complete one-loop renormalization of the vertices
$H_FZ^*\gamma^*$ and $H_F\gamma^*\gamma^*$ where the superscript $({}^*)$
means that the gauge boson may or may not be
real\refmark{\diazweiler,\BarrPR}.
The calculation is simplified if we expand the sum $\Sigma_{HV_1V_2}$
of all Feynman diagrams
contributing to the vertex $HV_1V_2$, $V_i=\gamma$ or $Z$,
in the following form factors:
$$\Sigma_{HV_1V_2}^{\mu\nu}(p^2,k^2,k'^2)=A_{V_1V_2}g^{\mu\nu}+
B_{V_1V_2}k^{\nu}k'^{\mu}+...\eqn\formf$$
where $p^2$ is the momentum squared of the Higgs, and $k^{\mu}$
($k'^{\nu}$) is the four-momentum of the vector boson $V_1$ ($V_2$).
There are three other form
factors but they do not contribute to the total widths we calculate.
If $V_1$ or $V_2$ is a real $\gamma$, then gauge invariance imposes
the following relation between the two relevant form factors:
$$A_{V\gamma}=-(k\cdot k')B_{V\gamma}\eqn\GaugeRel$$

\FIG\HZgamma{Contributions to the renormalized vertices $HZ\gamma$
and $H\gamma\gamma$. In the latter case, only the vertex corrections
contribute, and the vertex counterterm is zero. In the former case,
a non-zero vertex counterterm is necessary, as well as the graphs mixing
the $Z$ and the photon. Mixing between the Goldston boson $G^0$ and the
photon does not contribute to the relevant form factors $A_{Z\gamma}$
and $B_{Z\gamma}$.}
\FIG\IrrTriaD{Irreducible triangular diagrams contributing to the
one-loop induced veteces $HZ^*\gamma^*$ and $H\gamma^*\gamma^*$, where
the superscript $({}^*)$ means that the gauge boson may or may not be
on-shell. Charged Higgs loops are small, model-dependent, and
omitted.}
\FIG\IrrBubD{Irreducible bubble diagrams contributing to the one-loop
induced vertices $HZ^*\gamma^*$ and $H\gamma^*\gamma^*$.}

There are no tree level vertices $H\gamma\gamma$ or $HZ\gamma$;
these are induced at the one-loop level. In Fig.~\HZgamma\ we show the
different contributions to the two vertices.
There are two kind of diagrams that contribute to the vertex
$HZ^*\gamma^*$: the irreducible and the reducible. The latter
diagrams involve the mixing between the $Z$ gauge boson and the
photon. Both groups of diagrams are infinite and each one needs a
counterterm. The mixing between the Goldstone boson $G^0$ and the
photon does not contribute to the form factors $A$ and $B$.
In opposition to the previous case, the sum of the irreducible
diagrams contributing to the vertex $H\gamma^*\gamma^*$
is finite and the counterterm is zero. There is no contribution from
reducible diagrams to $H\gamma^*\gamma^*$. In Fig.~\IrrTriaD\
and~\IrrBubD\ we display the irreducible diagrams contributing to the
two vertices $HZ^*\gamma^*$ and $H\gamma^*\gamma^*$. They can be divided
into triangular diagrams (Fig.~\IrrTriaD) and non-triangular
(Fig.~\IrrBubD), induced by four point vertices involving four gauge
bosons or two scalars and two gauge bosons. We neglect charged scalar
loop contributions, since scalar loops are much smaller (and model
dependent).

\REF\diazHud{Marco A. D\'\i az, {\it Phys. Rev. D} {\bf 48}, 2152 (1993).}
\REF\Kniehl{B.A. Kniehl, {\it Nucl. Phys.} {\bf B376}, 3 (1992).}

The renormalization scheme used in the gauge boson sector
is outlined in ref.~[\diazHud] (see also ref.~[\Kniehl]): the
physical masses of the $Z$ and $W$ gauge bosons are given by the pole
of each propagator, the tree level $\gamma e\bar e$ vertex is fixed by the
electric charge measured at $q^2=0$, and no mixing
between the photon and the $Z$ gauge boson at $q^2=0$ is imposed.

\FIG\brsm{Branching ratios of the Standard Model Higgs.}

\REF\HtoGauB{Standard Model branching ratios, including one-loop
renormalization, can be found in:
D.Yu. Bardin, P.Ch. Christova, and B.M. Vilensky, report
number JINR-P2-91-140.}

\FIG\brphi{Branching ratios of a fermiophobic Higgs. The decay modes
$W^*W^*$ and $Z^*Z^*$ are calculated at tree level. All the others
include one-loop corrections to the vertices $HZ\gamma$ and
$H\gamma\gamma$. The decay mode $\gamma b\bar b$ includes also
box diagrams in order to ensure gauge invariance. $W^*$ and $Z^*$
decays are summed over all fermion final states.}

For comparison, we start in Fig.~\brsm\ showing some branching ratios of
the Standard Model Higgs\refmark\HtoGauB. We see that the dominant
decay mode is $H\rightarrow b\bar b$ with a branching ratio equal to
unity up to a Higgs mass approximately equal to $150$ GeV. Above that
mass, $H\rightarrow W^*W^*$ becomes dominant. The $H\rightarrow
\gamma\gamma$ branching ratio lies between $10^{-4}$ and $10^{-3}$.
In Fig.~\brphi\ we plot the branching ratios of the different decay modes
of the fermiophobic Higgs.
We focus our attention on the following decay modes:
$W^*W^*$, $Z^*Z^*$, $ZZ$, $Z\gamma$, $\gamma\gamma$, $Zb\bar b$,
$\gamma b\bar b$, $Z\Upsilon$, $\gamma\Upsilon$, and $\Upsilon b\bar b$.
We also explore other vector mesons like $\rho$ and $J/\Psi$. In the
modes $Z\gamma$, $\gamma\gamma$, and $\gamma b\bar b$, the gauge invariance
constraint of eq.~\GaugeRel\ is checked algebraically and
numerically. In the latter case, it is necessary
to include box diagrams in order to maintain gauge invariance. We do so.
The decays into two gauge bosons $W^*W^*$ and $Z^*Z^*$
are treated at tree level. In a forthcoming publication\refmark\diazweiler
we will include the loop induced $\gamma^*\gamma^*$ and $Z^*\gamma^*$
contributions as well. The virtual gauge boson decays into two fermions and
we sum over all possible fermions in the final state. The rest of the
decay modes include one loop radiative corrections through the vertices
$HZ\gamma$ and $H\gamma\gamma$. For example, in the decay $H_F\rightarrow
Zb\bar b$ there is a contribution from the tree level vertex $HZZ^*$
with one of the $Z$ bosons being off-shell and decaying into a pair
$b\bar b$, and a contribution from the one-loop induced vertex
$HZ\gamma^*$ with the off-shell photon decaying into $b\bar b$.

The decay mode $H_F\rightarrow\gamma\gamma$ is dominant if the Higgs
mass is less than about $90$ GeV. For Higgs masses above $100$ GeV the
main decay mode is $W^*W^*$ followed by $Z^*Z^*$. The ratio
between the tree level $Z^*Z^*$ branching ratio and
the branching ratio $Zb\bar b$ calculated also at tree level (including
only the tree level vertex $HZZ$) is $B(Z\rightarrow b\bar b)
\approx 0.15$. However, here we compare the tree level $Z^*Z^*$
mode with the one-loop corrected $Zb\bar b$ mode. This permits us to
appreciate the effect of radiative corrections on the decay
$H_F\rightarrow Zb\bar b$: the loop-induced $Z\gamma^*$ amplitude
exceeds the tree-level $ZZ^*$ amplitude for this case.

\FIG\brphii{Branching ratio involving a vector meson: $ZM$,
$b\bar bM$ and $\gamma M$, where $M$ may be $\rho$, $J/\Psi$ or
$\Upsilon$.}

\REF\mesons{M.N. Doroshenko, V.G. Kartvelishvili, E.G. Chikovani, and
Sh.M. \'Esakiya, {\it Sov. J. Nucl. Phys.} {\bf 46}, 493 (1987);
V. Barger, K. Cheung, and W.-Y. Keung, {\it Phys. Rev. D} {\bf 41},
1541 (1990).}

We also find that the decay modes involving the vector
meson\refmark\mesons
$\Upsilon$ have branching ratios as high as $10^{-5}$. Above $100$ GeV
the most important Upsilon mode is $Z\Upsilon$ and below that mass,
$\gamma\Upsilon$. Other vector mesons are studied in
Fig.~\brphii\ where we make a comparison between $\rho$, $J/\Psi$ and
$\Upsilon$. For the light mesons, the decay $H_F\rightarrow b\bar bM$
($M=\rho$ or $J/\Psi$) becomes dominant over the $ZM$ modes
for large Higgs masses. The $\rho b\bar b$
($J/\Psi b\bar b$) mode may be as high as $10^{-2}$ ($10^{-4}$).
Below $100$ GeV the most important mode is $\gamma M$ in the three
cases studied ($M=\rho$, $J/\Psi$ or $\Upsilon$), and for the light meson
$\rho$ the branching ratio is about $10^{-3}$.

\FIG\uppick{ Signal for a fermiophobic Higgs decaying into $Z^*\Upsilon$
compared with the $Z^*Z^*$ background (calculated at tree level).}

A clean signature is provided when $\Upsilon$ or $J/\Psi$
decays into a pair of charged leptons. As an example, we
compare the decay mode $H_F\rightarrow Z^*\Upsilon$ to the background
$H_F\rightarrow Z^*Z^*$ in Fig.~\uppick. Although the Upsilon production
rate is small, the $\Upsilon$ stands out over the background
(in this figure we present tree level calculations only).

In summary, the main discovery channel of a light
fermiophobic Higgs ($m_H\lsim 90$ GeV) is $H_F\rightarrow\gamma\gamma$
with a branching ratio close to the unity. The
production mechanism at LEP will be suppressed by the mixing factor
$\cos^2\beta=v_F^2/\sum v_i^2$, where $v_i$ is the vev of the ith Higgs
and $v_F$ is the particular vev of the fermiophobic Higgs
(in opposition to hadron machines, where the gluon-gluon/top-loop
production will be suppressed by fermiophobia). Above 90 GeV, decay is
dominantly to four fermions/jets, but modes with $Z$, $\gamma$, and
vector mesons like $\Upsilon$, $J/\Psi$, and $\rho$ are possibly
useful. Vector meson branching ratios are larger for lighter mesons.
With production of the fermiophobic Higgs at LEP suppressed by the
$\cos^2\beta$ factor, the fermiophobic Higgs could be light.

\vskip .5cm
\centerline{\bf ACKNOWLEDGMENTS}
\vskip .5cm

This work was supported by the U.S. Department of Energy, grant No.
DE-FG05-8SER40226.

\refout
\figout
\end